# Peaks in the electronic density of states are not strong predictors of Superconductivity

Zoë S. Yang[†,+], Austin M. Ferrenti[†], and Robert J. Cava[†,*]

[†] *Department of Chemistry, Princeton University Princeton NJ 08540 USA*

Electronic bands near the Fermi energy with minimal energy dispersion in $k$-space, i.e. "flat bands", are often said to be an important characteristic of superconducting materials. When extending over a significant region of the $k$-space, this leads to peaks in the electronic density of states. If one could systematically search the calculated electronic structures of materials, already available through the Materials Genome Initiative, for such peaks, then the rate of new superconductor discovery could be accelerated greatly. Here we test whether this characteristic actually distinguishes known superconductors from known non-superconductors. We find that there is an anticorrelation for tetragonal symmetry materials, but that hexagonal symmetry superconductors exhibit such peaks more frequently than hexagonal non-superconductors. More generally, our study's limitations underscore the importance of standardizing digital resources for achieving the accelerated materials discovery paradigm promised by the Materials Genome Initiative.

[+]Current Address: Citrine Informatics, 2629 Broadway St, Redwood City, CA 94063

**Introduction**

We test a specific hypothesis about the occurrence of superconductivity in materials. Our experiment, performed on the Materials Genome database, is designed to determine whether a specific electronic characteristic, "flat bands", manifested as peaks in a material's electronic density of states (DOS), are found more frequently in known superconductors than in known non-superconductors. To perform this experiment, we first created a machine-searchable database, compiled from different sources, of the formulas, crystal structures and electronic properties of 17,000 materials. We then searched their calculated electronic structures in the Materials Genome database, where available (a smaller, but still significant set), for this specific characteristic. Our results indicate that this electronic property is seen more often in hexagonal symmetry superconductors than in hexagonal symmetry non-superconductors, but we conclude that our study is limited by the availability, quality, and organization of existing data sets, indicating that standardization of machine-searchable databases will be required to take full advantage of the information made available via the Materials Genome Initiative.

If bands with minimal energy vs. electron $k$-vector variation near the Fermi energy ($E_F$) in the electronic structure of a solid, i.e. flat bands (FBs), which lead to peaks in the DOS, are important conditions for the occurrence of superconductivity, and if there was a method to automate the detection of these electronic structure features, then researchers could data mine for potential superconductors by screening calculated electronic structures. This paper makes use of computational tools, from the perspective of data mining and Big Data, to examine this possibility.

Although the relationship between flat bands and superconductivity is frequently cited in materials physics, two specific theories that attribute a flat band structure or high density of states near the Fermi energy of solids to superconductivity are the Flat/Steep Band and Van Hove Scenarios. The Van Hove Scenario relates the property of a Van Hove singularity (VHs), described as a saddle point in a material's electronic band structure, to superconductivity[1,2]. The lesser known, yet similar Flat/Steep Band Scenario argues that materials with flat bands near $E_F$, accompanied by "steep" bands crossing $E_F$, characterize superconducting materials[3,4]. Some argue that FBs near $E_F$ are a necessary condition for superconductivity[1,3,5]. Regardless of the scenario, the existence of FBs or VHs's in the band structure of a solid, when existing over a significant region of *k*-space, result in local maxima in the DOS (DOS = The Density of States - the number of electron energy states available within a narrow energy window, as a function of electron energy) at $E_F$[6,7]. As a result, a programmed search for peaks near $E_F$ in the output of DFT (Density Functional Theory) DOS calculations should allow for the detection of these electronic structure features without human intervention.

While DFT calculations can be time and computation expensive, the current age of digital materials has increased access to electronic structure property data in online materials databases. This offers an opportunity to test the hypothesis that peaks in the electronic density of states lead to superconductivity on a massive scale. Theoretically, if a data-mining type database search found that peaks in the DOS near $E_F$ appeared more commonly in superconductors than in non-superconductors, then the existence of such peaks near $E_F$ could be used as a computation-based screening parameter to narrow the search for new superconductors.

However, superconductivity research to date has demonstrated many challenges in working with data. Recent data-related research concerning superconductivity [8–11] all seems to express a common grievance of insufficient or unreliable data. The use of non-uniform or highly clustered data, for instance, could impede the accuracy of AI[11] and general data science research. Furthermore, researchers have noted difficulty with missing but important information about superconductors (crystal structure, for example) in the available databases[8,9]. Within such limitations, we created a database of known superconductors, and executed a search on the on-line accessible Materials Project DOS data to compare the frequency of maxima near $E_F$ in superconductors and non-superconductors, organized by crystal system. Our aim was to understand if and when FBs and their associated peaks in the DOS are unique to and/or concurrent with superconductivity.

**Results and Discussion**

For materials with calculated band gaps less than or equal to 0.1 eV, a maximum band gap taken to indicate the presence of a calculated metallic or dopable state, our search detected DOS peaks within 0.1 eV of $E_F$ in 33% more of the species in our database of hexagonal superconductors than in our equally-sized database of hexagonal non-superconductors. To this extent, our results suggest that metals in the hexagonal crystal system possess FBs and DOS peaks near their Fermi energies more commonly in superconducting materials than in non-superconducting materials.

As it is not known, independent of a theoretical model, how close to the Fermi energy DOS peaks would need to be to give rise to superconductivity, our experiments were performed to gradually narrow in on the Fermi Energy when looking for such peaks. We find that the percentage of superconductors exhibiting DOS peaks decreased monotonically in this narrowing. Within 1 eV of the Fermi energy, 472, or 51.9% of the superconductors had at least one local maximum in their DOS. At 0.75, 0.5, 0.25, and 0.1 eV from the Fermi energy, the percentage of superconductors in our data demonstrating peaks decreased from 41.3%, to 28.1%, to 14.7%, to 7.4%, respectively. A similar decrease in the prevalence of DOS peaks occurred when closing in on $E_F$ for the non-superconductors, making it seem as though a DOS peak near the Fermi energy is not particularly unique to superconductors.

In spite of similar trends in the frequency of DOS peaks near $E_F$ for superconductors and non-superconductors, our data mining experiment does demonstrate differences in the relative ratios of superconductors containing DOS peaks near $E_F$ to the non-superconductors with DOS peaks. The number of superconductors with DOS peaks for the hexagonal crystal system, for example, exceeded the upper bound of the standard error for non-superconductors with DOS peaks within 0.1 eV from $E_F$. The data mining experiment showed that hexagonal materials with local maxima within 0.1 eV of $E_F$ were 1.33 times more numerous among superconductors than among non-superconductors. Accordingly, our experiment implies that that the existence of a peak in the DOS of hexagonal materials may be more auspicious in the search for superconductors than in other crystal systems.

Contrary to the case of hexagonal superconductors, the percentage of cubic and tetragonal superconductors with DOS peaks near $E_F$ was *smaller* relative to the percentage of non-superconductors with DOS peaks in the same energy range. This difference was particularly dramatic for the tetragonal crystal system. Within 0.1 eV from the Fermi energy, the mean fraction of non-superconductors with peaks was 3.07 times as large as the fraction of superconductors found to have peaks.

For the orthorhombic and trigonal crystal systems, the incidence of peaks in the DOS for superconductors and non-superconductors did not vary significantly relative to one another. Perhaps orthorhombic and trigonal polymorphs of superconductors are no less likely to contain peaks in the DOS than orthorhombic and trigonal polymorphs of non-superconductors. In this case, a local maximum in the DOS near the $E_F$ of orthorhombic or trigonal structures with small or negligible calculated band gaps might have little predictive power, at least on its own, for identifying potential superconductors. The main results of this study are summarized in Tables 1 and 2 and Figure 1.

Although these results seem unpromising outside of the hexagonal case, our data mining experiment was limited by both the need to apply a Gaussian smear to the calculated DOS and the omission of spin orbit coupling from existing calculations. The limitations point to a potentially broader issue in employing data-based searches for materials with particular physical properties. The varying resolutions of the DOS calculations currently available on the Materials Project database made it necessary to apply a Gaussian broadening to the DOS data, with a relatively large standard deviation of $\sigma = 0.5$, using pymatgen[a]. In some cases, the smearing adequately smoothed out graphs that were too choppy, but in others, peaks were lost or shifted as they were averaged into neighboring peaks.

Comparison of DOS curves obtained from the Materials Project with $\sigma = 0.5$ and $\sigma = 0.1$ is illustrative. In general, the large Gaussian filter caused very small peaks near $E_F$ to not get detected when they were very close to larger peaks but did not always result in false negatives. For $Ba_4Si_{23}$, for example (Figure 2), the tiny peaks near $E_F$ present with the smaller Gaussian filter got averaged into a broader curve, but the code still detected a local maximum. Sometimes the low resolution of the DOS data itself also hindered the analysis. Comparing the DOS of $La_2CuO_4$ obtained from the Materials Project to the DOS for $La_2CuO_4$ recalculated from the input files on the Materials Project database with more data points, for example, demonstrates a significant absence of information in the DOS curve obtained from the Materials Project for the purposes of this data mining experiment (Figure 3). While the overall shape of the Materials Project DOS data for $La_2CuO_4$ matches the overall shape of both a published DOS of $La_2CuO_4$ [12] and our recalculated DOS, the Materials Project data plot fails to exhibit a peak at its Fermi energy. Calculations do not always include enough points in the DOS to capture small, sharp peaks like the one in $La_2CuO_4$ [13]. In fact, it is not uncommon for DOS calculations to have resolutions too low to show sharp peaks in high $T_c$ cuprates, and in such cases researchers have suggested that full E vs. $k$ band structures may provide a better picture of VHs's [14]. Thus, as was the case for $La_2CuO_4$, our experiment may have missed DOS peaks within 1 eV from $E_F$ for cuprates and other materials with small peaks that are missing from lower resolution DOS data.

Finally, the Materials Project calculations do not take spin-orbit coupling into account[15]. The recalculated DOS's of $KBi_2$ from the Materials Project with and without spin-orbit coupling demonstrate how that factor can affect the shape of the DOS (Figure 4) for materials based on heavy elements. When plotted with $\sigma = 0.1$, the DOS peak nearest to $E_F$ shifts slightly closer to $E_F$, but the peaks closest to it shift away from $E_F$ when spin-orbit coupling is included in the calculation. When plotted with $\sigma = 0.5$, the inclusion of spin-orbit coupling makes the DOS peak near $E_F$ shift in the positive direction. Although our code detected peaks near $E_F$ for the DOS plots of $KBi_2$ with and without spin-orbit coupling, in other instances, omitting spin-orbit coupling could cause peaks to appear in inaccurate locations. For this reason, the detection of DOS peaks for superconductors containing heavy elements may not have been as effective as it was for the general case.

**Conclusions**

Our findings concerning hexagonal superconductors suggest that performing a mass search on databases for hexagonal symmetry metals with DOS peaks within 0.1 eV of their Fermi energies, and subsequent synthesis and testing of these materials for superconductivity, would be interesting follow-up experiments. For tetragonal materials, peaks in the DOS do not correlate with superconductivity, and we speculate that such peaks for tetragonal symmetry materials may lead to competing physical properties over broad ranges of materials types.

Thus, we conclude that the correlation between superconductivity and peaks in the DOS and the flat bands that lead to them, when tested against a broad range of known materials, is spotty at best, clear only in the hexagonal crystal system, contrary to commonly held views. More generally, however, the challenges we faced throughout our data mining experiment demonstrate the restrictions that the current availability and organization of materials data place on reliable machine-learning-based experimentation. We had to code many extra scripts, for example, in order to compare chemical formulas written using incompatible conventions in different databases. With no easy way to compare the identities of entries across sources, the original 17,375 entries in our database was narrowed to only 980 superconductors with available computed DOS data. Ideally, we would have been able to easily cross-check materials' identities among databases, so that we could reliably pull data from multiple sources without writing extra code. Thus, we advocate for further work to standardize the organization of materials databases, so that entries better normalize historical, field, or culture dependent idiosyncrasies in naming conventions, units, and exceptions.

Furthermore, the quality of the DFT-calculated DOS's could be standardized. To reach more impactful conclusions, not only in this experiment, but also in Machine Learning experiments including DOS data in general, at the present time scientists must rely on their own DFT calculations with more data points. As a result, materials databases could better serve data-science research pertaining to electronic properties by striving for higher quality calculations for all entries. Thus, although we conclude that data science has the possibility to revolutionize the way that researchers approach materials discovery, we believe the Materials Genome Initiative's promise of discovery acceleration cannot be fully advantaged without further database and computational development.

**Methods**

We carried out this research in three distinct phases. The first phase involved creating the necessary technical frameworks to handle the data used in subsequent experiments. We created our own database, and populated it with information from online materials databases. The second phase concerned developing and testing a program to detect FBs. Finally, we compared the incidence of FBs in superconductors to their incidence in non-superconductors.

**Data Management and Acquisition**

To simplify the process of working with multiple different data sources, each with its own query syntax, API, and data format, we created an internal relational database of superconductors and their properties. We managed the database through a simple querying API we developed using Python's sqlite3 library [16], along with the SQLite Studio [17] interface for manual input. Since materials databases tend to list materials under differing chemical naming conventions, we used pymatgen's Composition class, along with our own regular expression and formula-parsing functions, to convert all formulas so that they followed the same protocols. The use of our own superconductor database also allowed us to consolidate data with different units and different formats, like data structures, json, or csv, in one place under the same unit standards. Furthermore, it separated the data collection process from the experimentation process, and eliminated the need to constantly make requests to remote servers through multiple unlike querying API's at once.

The database was populated with superconductors from three major sources. We pooled superconductor data and references from two public data sets on Citrine Informatic's Citrination platform using their Pypif API [18]. One set was compiled by Citrine Informatics, and contained superconductors and their critical temperatures (data set 2210) [19]. The other was the National Institute of Standards and Technology (NIST) list of high temperature superconductors (data set 151804) [20].

These two data sets from Citrine's platform made up 1,070 distinct superconductors (by composition or crystal system) total in our database. 15,986 more superconductors were added to the database from Valentin Stanev's github repository [21] linked in Stanev et al.'s "Machine learning modeling of superconducting critical temperature" [8]. Stanev et al. originally curated this data from the National Institute for Materials Science's SuperCon database [22]. Finally, 318 entries from B. W. Robert's "Survey of superconductive materials and critical evaluation of selected properties." [23] were added manually. We included references for each data point in our database, and included critical temperatures and crystal systems for the superconductors when available.

We also tied data from the Materials Project to the unique entries in our superconductor database. We selected formulas and existing crystal structure data from our database, and queried the Materials Project database for species with the same IUPAC reduced formulas and crystal structures. We then added the species' corresponding Materials Project ids to our database. If entries in our database did not have crystal structure information, we assigned Materials Project ids to entries in our database only if one polymorph existed for their IUPAC reduced formulas in the Materials Project database. We added more ids based on the articles referenced in the sources from which the superconductors originated. The Materials Project ids were saved for the purposes of querying for electronic data that would require too much memory to save on a local computer, as well as to populate our database with more crystal structure information.

**Detecting Flat Bands**

We programmatically detected FBs in superconductors' electronic structures by applying a function implementing the first and second derivative tests to DOS data to identify maxima near the superconductors' Fermi energies. Documentation for the function, called findVHS(), can be found in

Table S1. The code takes the energies and densities pertaining to a given DOS, and returns the energy and density corresponding to a local maximum in the DOS within a specified energy range from the Fermi energy, should such a peak exist.

The parameters to findVHS(), "dVals," "eVals," "index range," "deriv1_tol," and "deriv2_tol," were input as the python data structures of a numpy array, another numpy array, a float, a float, and another float, respectively. "dVals" contained the density values for the DOS, and "eVals" contained each of the corresponding energy values minus the Fermi energy. The "index bounds" parameter designated the distance from the Fermi energy that should be searched for a peak, so that findVHS() could more efficiently identify local maxima. Furthermore, a binary search algorithm[24] was used to cut down the time required to isolate the energies and densities pertaining to the index range.

Once findVHS() obtained the desired energies and densities index bounds from $E_F$, it applied the first and second derivative tests. However, a few measures were taken to increase the flexibility of findVHS() when dealing with imperfect data. The tolerance parameters, deriv1_tol, and deriv2_tol, allowed the client to set a tolerance on the value of zero the program considered 0 for the first derivative test (i.e. anything within 0.05 eV of 0 was considered 0). deriv2_tol allowed the user to set the largest value findVHS() considered as negative for the second derivative test. To ensure that the tolerance parameters affected all DOS plots in the same manner, the derivatives of the DOS plots were taken with the densities normalized by the maximum density corresponding to the energy range specified by index bounds.

Due to variations in the DOS data that we obtained from the Materials Project, we used pymatgen to apply a Gaussian smear/filter, with a maximum standard deviation of $\sigma = 0.5$, across all density arrays obtained. The maximum standard deviation of $\sigma = 0.5$ was chosen, because lower values of $\sigma$ did not sufficiently smooth the DOS data from the Materials Project for all the plots. deriv1_tol, and deriv2_tol were calibrated to 0.15, and 0, respectively. This was accomplished by tweaking deriv1_tol and deriv2_tol, then evaluating the code for accuracy graphically. More specifically, the DOS's used to test the code were plotted along with the graph of y = E, where E was the energy location of the peak in the DOS within 1 eV of the Fermi energy returned by findVHS(), should such a peak exist. Visual inspection for the intersection of y = E and a maximum in the DOS near $E_F$, or lack thereof, exposed the likelihood of false negatives or false positives in the results.

To assess potential errors due to the Gaussian filter, we also plotted some of the DOS curves with $\sigma = 0.5$ on the same plot with $\sigma = 0.1$. To further assess the reliability of findVHS(), and explain its limitations, we performed DFT calculations to obtain the DOS of $La_2CuO_4$, as well as the DOS of $KBi_2$ with spin-orbit coupling (SOC). DFT calculations were carried out using the Vienna ab initio Software Package (VASP) [25-28] using the Projector Augmented Wave Method and the Linear Tetrahedron Method for integration over the Brillouin Zone[29]. Pymatgen's pymatgen.io.vasp [30] module was used to parse the output of the VASP calculations, and obtain the final DOS data. The results were plotted and compared to the DOS plots for $La_2CuO_4$ and $KBi_2$ obtained from the Materials Project database with different values of $\sigma$ for the Gaussian filter.

FindVHS() was ultimately run on all available DOS data from the Materials Project for the superconductors in our database, and the existence (or non-existence) of a peak saved. The number of superconductors with DOS peaks within 1, 0.75, 0.5, 0.25, and 0.1 eV from $E_F$ was recorded for future comparison with the number of DOS peaks within these energy bounds in non-superconductors.

**Evaluating the Incidence of FB's in Superconductors vs. Non-Superconductors**

To assess the incidence of FBs in superconductors compared to their incidence in non-superconductors, we requested sets of "non-superconductors" from the Materials Project, and searched these sets for FBs. Selection statements to our database superconductors revealed our database predominantly comprised superconductors with metallic band gaps (≤0.1 eV), so we used the MPRest API[37] to query for "non-superconductors" with band gaps within the metallic range. To reduce the chance of accidentally requesting superconductors, queries to the Materials Project were made to exclude anything that could be an A-15 phase, elemental species, or member of families containing Cu-O, O-W, B-Mg, B-C-O-Yb, and Ba-Bi-K-O.

For each representative crystal structure in our superconductor database, we queried the Materials Project database for all species with DOS data, and the aforementioned band gap and chemical family constraints. Until we had selected as many metallic non-superconductors as our sample size of metallic superconductors of the same crystal system, we ran a routine that started by randomly generating a number, r, in the domain of 0 to the size of the non-superconductor set minus 1. If the reduced formula of the $r^{th}$ species in the non-superconductor query did not correspond to an entry in our superconductor database, and if the non-superconductor had not yet been encountered, the species was added to the current list of non-superconductors. Otherwise, a new random number was drawn.

Once the size of a set of non-superconductors was as large as the size of the set of superconductors with DOS data in the database for a given crystal system, findVHS() was run on the DOS data for the non-superconductors. The process of sampling for non-superconductors and running findVHS() was repeated 10 times. The means and standard deviations of the number of non-superconductors with local maxima within 1, 0.75, 0.5, 0.25, and 0.1 eV from $E_F$ were computed for all 10 trials. There were more than 10 times the number of non-superconductors returned from the Materials Project queries as superconductors in the database. Additionally, there was little overlap in the reduced formulas of species in the Materials Project queries with those in the superconductor database. Thus, the experiments did not extensively re-sample the same non-superconductors throughout the 10 trials for the different crystal systems.

After running this experiment for cubic, orthorhombic, tetragonal, and hexagonal crystal systems with metallic band gaps, the 10-trial experiment was repeated without respect to crystal system or band gap. The results of all non-superconductor trials were compared to the number of superconductors with local maxima in their DOS's within 1, 0.75, 0.5, 0.25, and 0.1 eV of $E_F$.

# References

1. Bok, J. & Bouvier, J. Superconductivity and the van hove scenario. *Journal of Superconductivity and Novel Magnetism* 25, 657–667 (2012). URL https://doi.org/10.1007/s10948-012-1434-3.

2. Markiewicz, R. A survey of the Van Hove scenario for high-tc superconductivity with special emphasis on pseudogaps and striped phases. *Journal of Physics and Chemistry of Solids* 58, 1179–1310 (1997). cond-mat/9611238.

3. Simon, A. Superconductivity and the periodic table: from elements to materials. *Philosophical Transactions of the Royal Society A: Mathematical, Physical and Engineering Sciences* 373, 20140192 (2015). URL https://royalsocietypublishing.org/doi/abs/ 10.1098/rsta.2014.0192. https://royalsocietypublishing.org/doi/pdf/10.1098/rsta.2014.0192.

4. Deng, S., Simon, A. & Kohler, J. A "flat/steep band" scenario in momentum space. *Journal of Superconductivity* 17, 227–231 (2004). URL https://doi.org/10.1023/B:JOSC.0000021247.78801.78.

5. Deng, S., Simon, A. & Khler, J. The origin of a flat band. *Journal of Solid State Chemistry* 176, 412 – 416 (2003). URL http://www.sciencedirect.com/science/article/pii/S0022459603002391. Special issue on The Impact of Theoretical Methods on Solid-State Chemistry.

6. Hoffmann, R. How chemistry and physics meet in the solid state 26, 846–878. URL http://doi.wiley.com/10.1002/anie.198708461.

7. Felser, C. *Superconductivity and Magnetoresistance in Colossal Magneto Resistance in quasitwo-dimensional systems: a chemical point of view*.

8. Stanev, V. *et al.* Machine learning modeling of superconducting critical temperature. *npj Computational Materials* 4, 29 (2018).

9. Konno, T. *et al.* Deep learning of superconductors I: estimation of critical temperature of superconductors toward the search for new materials. *CoRR* abs/1812.01995 (2018). URL http://arxiv.org/abs/1812.01995. 1812.01995.

10. Liu, Y. *et al.* Prediction of superconducting transition temperature using a machine-learning method. *MATERIALI IN TEHNOLOGIJE* 52, 639–643 (2018).

11. Meredig, B. *et al.* Can machine learning identify the next high-temperature superconductor?

examining extrapolation performance for materials discovery. *Mol. Syst. Des. Eng.* 3, 819–825 (2018). URL http://dx.doi.org/10.1039/C8ME00012C.

12. Subramoniam, G., Rao, R. S., Jaya, S. M. & Asokamani, R. Electronic structure of la2cuo4. *Pramana* 33, 673–675 (1989). URL https://doi.org/10.1007/BF02845692.

13. Nedos- vaspwiki. URL https://cms.mpi.univie.ac.at/wiki/index.php/ NEDOS.

14. Markiewicz, R. Van hove singularity and high-tc superconductivity vi: Interlayer coupling. *Physica C: Superconductivity* 177, 171 – 182 (1991). URL http://www.sciencedirect.com/science/article/pii/092145349190315P.

15. Jain, A., Hautier, G., Ong, S. P. & Persson, K. New opportunities for materials informatics: Resources and data mining techniques for uncovering hidden relationships (supplementary information). *Journal of Materials Research* 31, 977994 (2016).

16. sqlite3. URL https://docs.python.org/3/library/sqlite3.html.

17. Sqlite studio. URL https://sqlitestudio.pl.

18. Citrine Informatics. Pypif. URL http://www.citrine.io/pypif.

19. Citrine Informatics. Superconductor critical temperatures (2016).URL https://citrination.com/datasets/2210.

20. National Institute of Standards and Technology. Nist-high temperature superconductors (2017). URL https://citrination.com/datasets/151804.

21. Stanev, V. Supercon (2018). URL https://github.com/vstanev1/Supercon. Data used in Machine learning modeling of superconducting critical temperature paper. Downloaded from SuperCon database. The first column (name) gives the chemical composition of the compound in format element - weight (number). The second column gives the critical temperature of the compound. Materials without reported critical temperature have been assigned Tc=0.

22. National Institute for Materials Science (NIMS). Supercon. URL https://supercon.nims.go.jp/index_en.html.

23. Roberts, B. Survey of superconductive materials and critical evaluation of selected properties.*J. Phys. Chem. Ref. Data; (United States)* 5.

24. Sedgewick, R. & Wayne, K. *Algorithms* (Addison-Wesley, 2011), 4 edn.

25. Kresse, G. & Hafner, J. Ab initio molecular dynamics for liquid metals. *Phys. Rev. B* 47, 558–561 (1993). URL https://link.aps.org/doi/10.1103/PhysRevB.47.558.

26. Kresse, G. & Hafner, J. Ab initio molecular-dynamics simulation of the liquid-metal–amorphous-semiconductor transition in germanium. *Phys. Rev. B* 49, 14251–14269 (1994). URL https://link.aps.org/doi/10.1103/PhysRevB.49.14251.

27. Kresse, G. & Furthmller, J. Efficiency of ab-initio total energy calculations for metals and semiconductors using a plane-wave basis set. *Computational Materials Science* 6, 15 – 50 (1996). URL http://www.sciencedirect.com/science/article/pii/ 0927025696000080.

28. Kresse, G. & Furthmuller, J. Efficient iterative schemes for ab initio total-energy calculations¨ using a plane-wave basis set. *Phys. Rev. B* 54, 11169–11186 (1996). URL https://link. aps.org/doi/10.1103/PhysRevB.54.11169.

29. Kresse, G. & Joubert, D. From ultrasoft pseudopotentials to the projector augmented-wavemethod. *Phys. Rev. B* 59, 1758–1775 (1999). URL https://link.aps.org/doi/10. 1103/PhysRevB.59.1758.

30. Jain, A. *et al.* A high-throughput infrastructure for density functional theory calculations (2011).

**Acknowledgements**

This research was supported by the US Department of Energy Division of Basic Energy Sciences, grant DE-FG02-98ER45706, and by Princeton University.

**Competing Interests**

The authors declare no competing financial or non-financial interests.

**Data Availability**

The authors declare that the data supporting the findings of this study are available within the paper.

**Author Contributions**

Z.S.Y. and R.J.C. devised the main conceptual ideas and guiding principles for the project. Z.S.Y. performed the data screening and analysis. R.J.C. supervised the research. Z.S.Y. and A.M.F. wrote the manuscript with input from all authors.

**Correspondence and requests for materials** should be addressed to R.J.C. (email: rcava@princeton.edu).

**Table 1** Percentage ranges of metallic non-superconductors and percentages of metallic superconductors with local maxima in the DOS within 1 eV of $E_F$ by crystal system

|  |  | non-superconductors |  |  | superconductors |  |
|---|---|---|---|---|---|---|
| Crystal System | # species | mean # Peaks for 10 trials | stdev of # Peaks for 10 trials | % Peaks range | # Peaks | % Peaks |
| cubic | 341 | 195.5 | 5.54 | 55.70-58.96 | 192 | 56.30 |
| tetragonal | 219 | 114.5 | 7.37 | 48.92-55.65 | 119 | 54.34 |
| orthorhombic | 116 | 48.7 | 7.67 | 35.30-48.60 | 43 | 37.07 |
| trigonal | 29 | 11.2 | 3.49 | 26.59-50.65 | 10 | 34.48 |
| triclinic | 3 | 0.4 | 0.52 | - | 3 | 100 |
| monoclinic | 26 | 7.6 | 3.17 | 17.04-41.42 | 4 | 15.38 |
| hexagonal | 175 | 104.2 | 6.19 | 55.95-63.02 | 101 | 57.71 |
| **Distance from Fermi energy: 1 eV** | | | | | | |

**Table 2** Percentage ranges of metallic non-superconductors and percentages of metallic superconductors with local maxima in the DOS within 0.1 eV of $E_F$ by crystal system

|  |  | non-superconductors |  |  | superconductors |  |
|---|---|---|---|---|---|---|
| Crystal System | # species | mean # Peaks for 10 trials | stdev of # Peaks for 10 trials | % Peaks range | # Peaks | % Peaks |
| cubic | 341 | 31.4 | 4.74 | 7.82-10.60 | 24 | 7.04 |
| tetragonal | 219 | 21.5 | 3.21 | 8.35-11.28 | 7 | 3.20 |
| orthorhombic | 116 | 9.0 | 2.83 | 5.32-10.20 | 10 | 8.62 |
| trigonal | 29 | 1.4 | 0.84 | 1.92-7.74 | 1 | 3.35 |
| triclinic | 3 | 0 | 0 | - | 0 | 0 |
| monoclinic | 26 | 1.4 | 0.84 | 2.14-8.63 | 0 | 0 |
| hexagonal | 175 | 18.7 | 3.65 | 8.60-12.77 | 25 | 14.29 |
| **Distance from Fermi energy: 0.1 eV** | | | | | | |

**Figures**

**Figure 1** Percentage of superconducting and non-superconducting species exhibiting peaks within (a) 1.0 eV and (b) 0.1 eV of $E_F$ in the calculated DOS plots by crystal class.

**Figure 2** Sharp peaks in the DOS averaged into broad peaks due to the size of the Gaussian smear ($\sigma = 0.1$ vs. $\sigma = 0.5$) for $Ba_4Si_{23}$.

**Figure 3** DOS of $La_2CuO_4$ as obtained from the Materials Project (a), and recalculated DOS of $La_2CuO_4$ (b), without Gaussian smearing.

**Figure 4** DOS of $KBi_2$ recalculated from the Materials Project Database with $\sigma = 0.1$ without spin-orbit coupling (a) and with spin-orbit coupling (b), and DOS of $KBi_2$ recalculated from the Materials Project Database with $\sigma = 0.5$ without spin-orbit coupling (c), and with spin-orbit coupling (d).

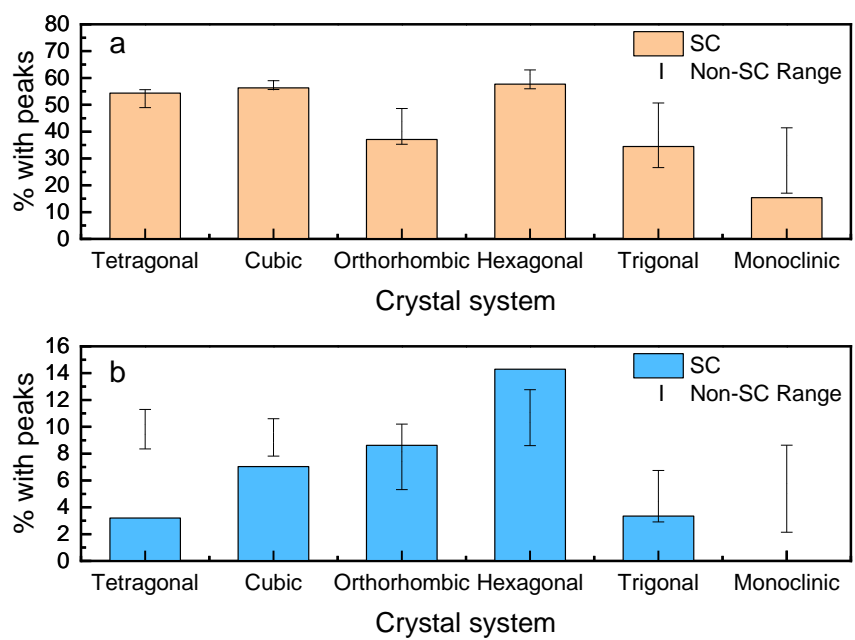

**Figure 1**

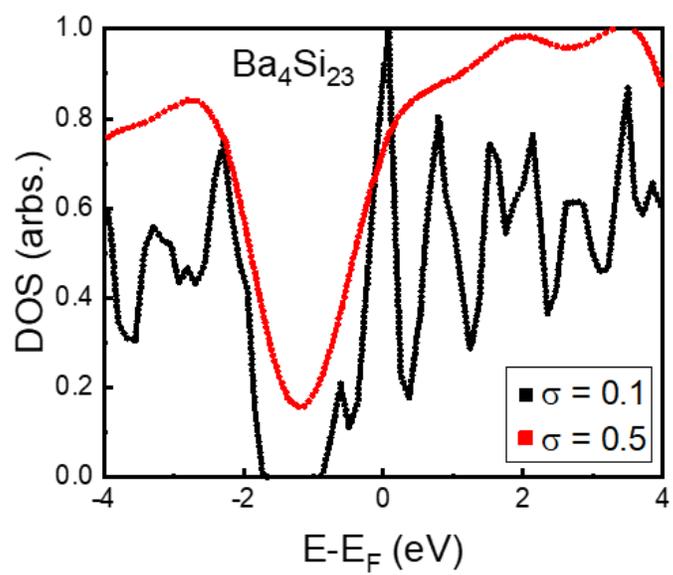

**Figure 2**

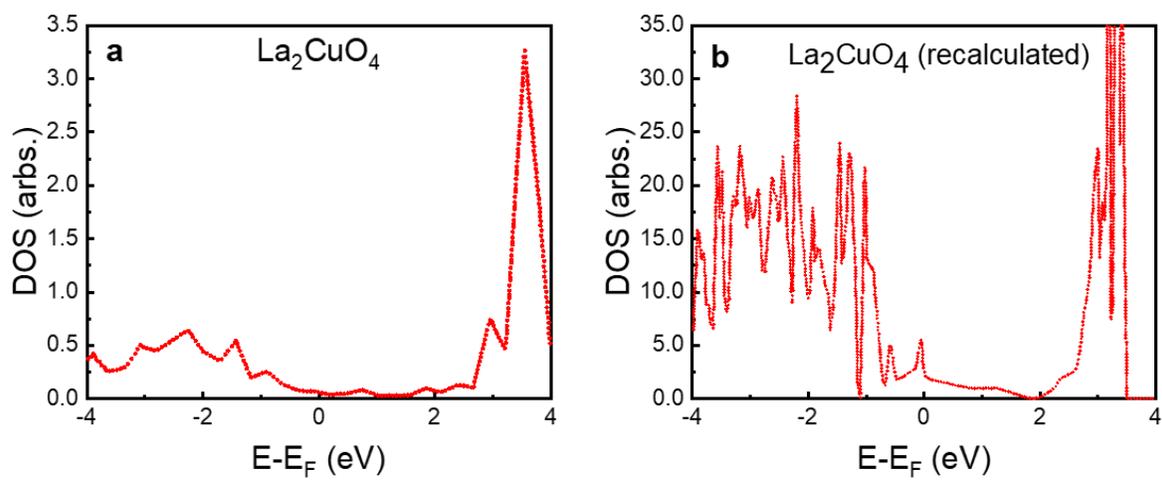

**Figure 3**

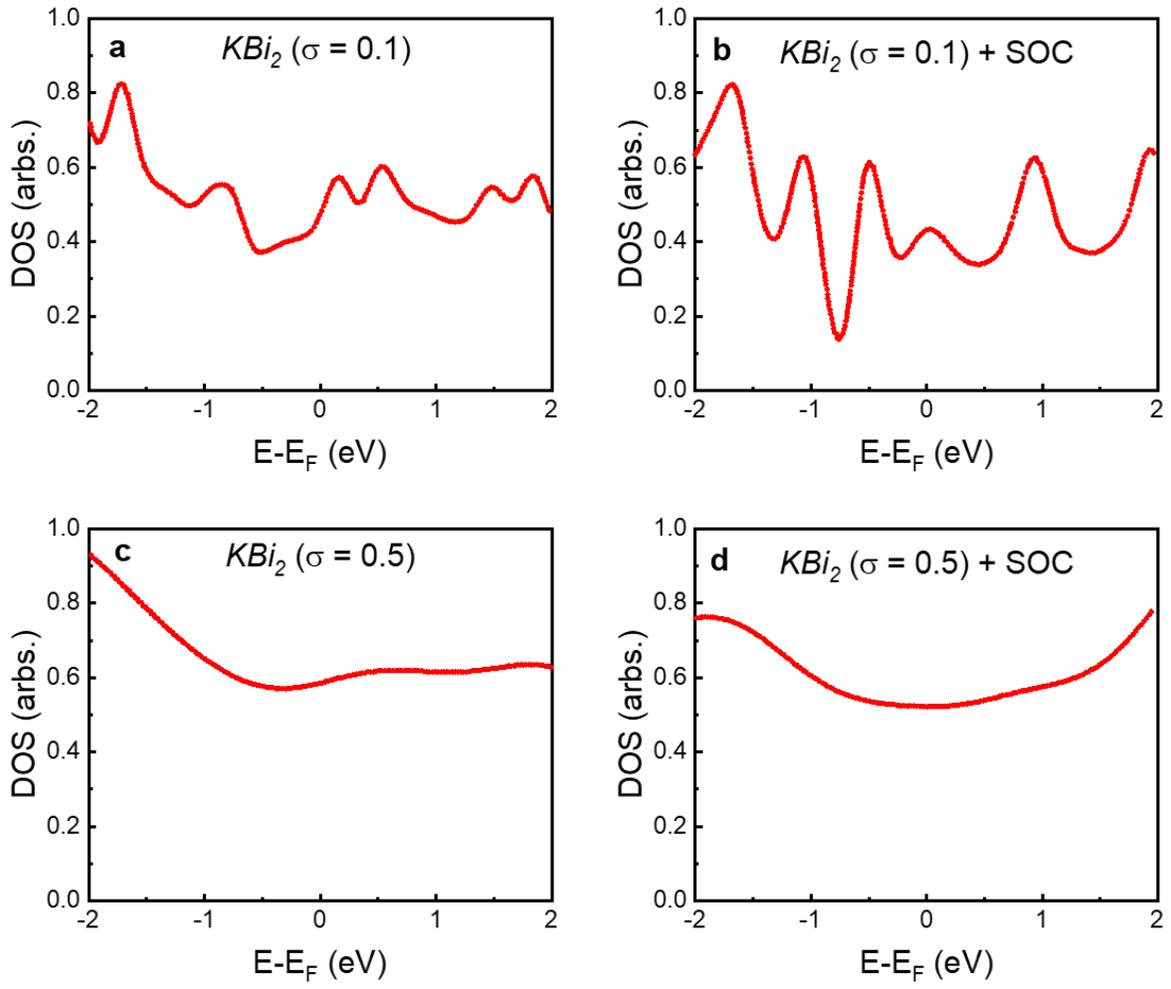

**Figure 4**